\documentclass[conference]{IEEEtran}
\IEEEoverridecommandlockouts
\usepackage{cite}
\usepackage{amsmath,amssymb,amsfonts}
\usepackage{algorithmic}
\usepackage{graphicx}
\usepackage{textcomp}
\usepackage{xcolor}
\usepackage{multirow}
\usepackage{booktabs}
\usepackage{makecell}
\usepackage{graphicx}
\usepackage{url} 
\def\BibTeX{{\rm B\kern-.05em{\sc i\kern-.025em b}\kern-.08em
    T\kern-.1667em\lower.7ex\hbox{E}\kern-.125emX}}
\begin{document}

\title{Activation Steering for Accent-Neutralized Zero-Shot Text-To-Speech
\thanks{This work was supported in part by Univ. of Texas at Dallas from the Distinguished University Chair in Telecommunications Engineering held by J.H.L. Hansen.}
}


\author{
\IEEEauthorblockN{Mu Yang, John H. L. Hansen}
\IEEEauthorblockA{
\textit{Center for Robust Speech Systems (CRSS), University of Texas at Dallas, USA} \\
\{mu.yang, john.hansen\}@utdallas.edu
}
}


\maketitle

\begin{abstract}
    Zero-shot Text-to-Speech (TTS) models can generate speech that captures both the voice timbre and accent of a reference speaker. 
    However, disentangling these attributes and individually controlling the output accent remains challenging. 
    In this study, we introduce a post-hoc and training-free approach to neutralize accent while preserving the speaker's original timbre, utilizing inference-time activation steering. We first extract layer-specific ``steering vectors" offline, which are derived from the activation differences within the TTS model between accented and native speech. During inference, the steering vectors guide the model to produce accent-neutralized, timbre-preserving speech. 
    Empirical results demonstrate that the proposed steering vectors effectively mitigate the output accent and exhibit strong generalizability to unseen accented speakers, offering a practical solution for accent-free voice cloning.
    We further confirm that speaker timbre is largely preserved via speaker embedding cosine similarity and speaker verification acceptance rate computed in an estimated accent-orthogonal subspace, as well as subjective listening tests. Audio demo: \url{https://accentsteer.github.io/}
\end{abstract}

\begin{IEEEkeywords}
Text-To-Speech, activation steering, accent neutralization, accent conversion
\end{IEEEkeywords}

\section{Introduction}

Zero-shot Text-To-Speech (TTS) is the task of generating speech for a target text using a reference speech utterance (and sometimes optionally, the reference speech's transcript) of an arbitrary speaker.
The generated speech is expected to have the voice characteristics of the reference speaker (e.g., timbre, prosody, accent, emotion, etc.). 
Recent advances on Diffusion Models and Large Language Models (LLM) have significantly improved zero-shot TTS \cite{chen2025f5,eskimez2024e2,Ju2024naturalspeech,kim2023p,zhu2025zipvoice, hu2026qwen3,wang2025spark,ye2025llasa, du2024cosyvoice1,du2024cosyvoice2,guo2025fireredtts,zhou2025indextts2,zhangvevo}. 

However, individually controlling different voice characteristics remains a challenge. 
When the reference speech has a second-language (L2) English accent, the generated speech often inherits both the accent and timbre from the reference. 
Although some prior instruction-based TTS models support dialect control, none of them support L2 English accent control in the zero-shot voice cloning scenario.
For example, CosyVoice3 \cite{du2025cosyvoice} only supports a fixed set of 26 instructions, none of which control the L2 English accent, and other instructions have minimal effects. Qwen3-TTS's \cite{hu2026qwen3} ``Voice Design” or ``Custom Voice Generation” modes (based on text prompts) cannot accept a reference speech sample to clone its timbre. 
In this work, we aim to generate speech with the timbre of the reference speaker but without the speaker's L2 accent. 
This is a practical problem for accent-free voice cloning, which can be useful for applications such as creating training targets for Accent Conversion models \cite{halychanskyi2025fac,quamer22_interspeech}, providing L2 language learners with personalized accent-neutralized speech feedback for computer-aided pronunciation training \cite{hirschi2025artificial}, etc.

Our solution is based on activation steering, a technique that modifies the internal activations of a neural network to control specific model behaviors.
It has been used to detoxify and shift the sentiment of LLM responses \cite{turner2023steering}, alter LLMs' persona traits \cite{chen2025persona}, and more \cite{li2023inference,rodriguez2025controlling,gaintseva2026casteer}.
The efficacy of this approach demonstrates that high-level semantic concepts can be represented as \emph{linear directions} in the activation space, and that steering the activations along these directions can effectively control the corresponding concepts in the model responses.
In this work, we ask the following research question: since the internal activations of a zero-shot TTS model encode multiple voice characteristics, can we steer the activations to neutralize the reference speaker's accent while preserving the timbre?

\begin{figure*}[t]
  \centering
  \includegraphics[width=0.95\linewidth]{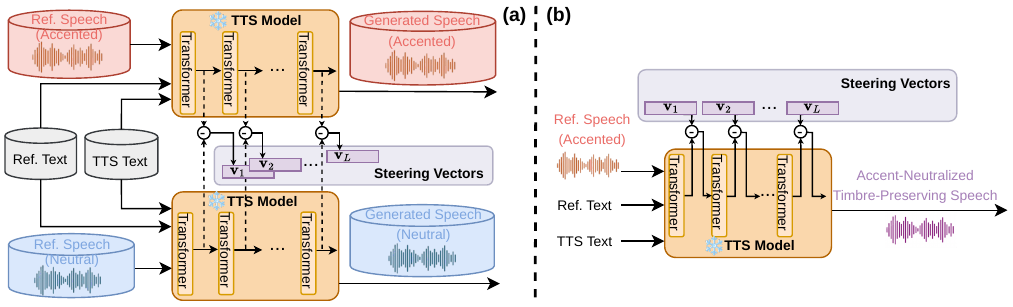}
  \vspace{-3mm}
  \caption{The proposed activation steering framework for accent-neutralized zero-shot TTS. (a) Steering vectors are extracted offline from the activation differences between accented and neutral speech, and (b) applied during inference to guide the model towards accent-neutralized output while preserving timbre. In this paper, we experiment with single-layer steering, i.e., only one layer is steered while other layers are left unchanged. The figure illustrates the general framework where multiple layers can be steered simultaneously.}
  \label{fig:system}
  \vspace{-4mm}
\end{figure*}

To this end, inspired by Persona Vectors \cite{chen2025persona}, we propose a \emph{post-hoc and training-free} approach. 
Specifically, as shown in Fig. \ref{fig:system}, we first extract ``steering vectors" offline from the layer-wise differences between the activation averages of accented and accent-neutral conditions.
Then, during inference, the steering vectors are applied to corresponding layers to guide the model to produce accent-neutralized, timbre-preserving speech.
We conduct steering experiments on Qwen3-TTS \cite{hu2026qwen3}, a state-of-the-art LLM-based zero-shot TTS model. Empirical results show that this simple steering approach can effectively mitigate the accent in the output speech, while maintaining the reference speaker's timbre to a large extent. The steering vectors also show strong generalizability to \emph{unseen} accented speakers,
whose utterances are not used for steering vector extraction,
suggesting that the steering vectors capture a general direction for accent neutralization in the TTS model's activation space. 
In addition, since raw speaker embedding cosine similarities may overstate timbre differences due to the intended accent change, we formulate objective measurements in an estimated accent-orthogonal subspace for a more faithful speaker timbre similarity evaluation.

\section{Related Work}

Several prior works have explored activation steering to enhance the controllability of zero-shot TTS. 
Suni et al. \cite{suni25_ssw} proposed a method to fit prosody and style vectors, which manipulate the embedding space of reference utterances. 
Counterfactual Activation Editing \cite{lee25f_interspeech} uses external classifiers and regressors to edit the internal representations of the TTS model to achieve post-hoc prosody and mispronunciation control. 
EmoShift \cite{zhou2026emoshift} builds lightweight activation steering layers on top of the output embeddings for emotion-aware TTS. 
TruS \cite{lee2026erasing} adopts a similar steering vectors extraction method for inference-time speaker unlearning in a Diffusion-based TTS model. 
Our work is related to EmoSteer \cite{xie2025emosteer}, which pre-extracts emotion steering vectors for a Diffusion-based TTS model and relies on an external emotion classifier to select top-k steering tokens, and thus requires multiple inference passes to apply the steering vectors. In contrast, our method does not require any external classifiers and
applies the steering vectors in a single autoregressive decoding pass, which is more efficient and practical for real-time applications.

\section{Method}

\subsection{The Qwen3-TTS Model}
We extract steering vectors from the activations of Qwen3-TTS \cite{hu2026qwen3}, a state-of-the-art LLM-based zero-shot TTS model. It uses a 12.5 Hz multi-codebook speech tokenizer with 1 semantic stream and 15 acoustic streams derived from Residual Vector Quantization (RVQ) \cite{defossez2024moshi}. The Qwen3-TTS model consists of a Qwen3-LLM-based backbone (28-layer Transformer) and a lightweight Multi-Token Prediction (MTP) module (5-layer Transformer) on top of the backbone. The backbone takes as input the aggregated codebook features from the speech tokenizer and predicts the semantic token, and the MTP module then predicts the remaining acoustic tokens. In this work, we focus on the activations from the backbone LLM, as it accounts for the majority of the model parameters and is responsible for learning the shared representations across different voice characteristics.

\subsection{Steering Vector Extraction} \label{sec:sv_extraction}
For zero-shot TTS, Qwen3-TTS takes as input a reference speech and its corresponding text transcript, and the target text to be synthesized. 
To compute the steering vectors, we use ARCTIC \cite{kominek2004cmu} and L2-ARCTIC \cite{zhao2018l2} to curate an extraction dataset with contrastive accent conditions.
ARCTIC provides US American English speech (considered as accent-neutral) from 4 native speakers, while L2-ARCTIC provides L2-accented English speech for the same set of sentences in ARCTIC, where each sentence is spoken by 4 L2 speakers. We randomly sample two disjoint sets of $M$ sentences, one as the target texts and one as the reference texts. For each target text, we pair it with one reference text and aggregate the corresponding speech from the 4 native speakers and 4 accented speakers, creating the (target text, reference text, reference speech) triplets as the TTS inputs.
As shown in Fig. \ref{fig:system}, we feed the triplets into the TTS model to generate accented and neutral speech. For each generated speech, we record each Transformer layer's output within the backbone LLM, averaged over the generated tokens. Then the steering vectors are computed as the difference between the mean activations of the accented condition and the neutral condition:
\begin{align}
    \mathbf{v}_l = \frac{1}{N_a} \sum_{i=1}^{N_a} \mathbf{a}_{l,i}^{(accented)} - \frac{1}{N_n} \sum_{i=1}^{N_n} \mathbf{a}_{l,i}^{(neutral)} 
    \label{eq:sv_extraction}
\end{align}
where $\mathbf{v}_l \in \mathbb{R}^{d}$ is the steering vector for layer $l$ with dimension $d$, $\mathbf{a}_{l,i}^{(accented)}$ and $\mathbf{a}_{l,i}^{(neutral)}$ are the averaged activations of the $i$-th generated sample at layer $l$, for the accented and neutral conditions, respectively, and $N_a$ and $N_n$ are the number of samples in each condition. We only keep track of the activations of generated tokens, while the activations of prompt tokens (i.e., reference speech and text) are excluded.

Since accent is coupled with speaker identity (i.e., the same speaker always speaks with the same accent), the steering vectors may capture not only the accent but also the speaker information. To break the entanglement and encourage the steering vectors to capture more accent-specific information, during steering vectors extraction, we apply on-the-fly data augmentations on reference speech waveforms to introduce perturbations that modify the speaker's voice \cite{qian2022contentvec,yang23v_interspeech}. These perturbations involve 3 sequential random transformations: 1) scaling all formant frequencies within an utterance by a random factor; 2) scaling the fundamental frequency (F0) by a random factor; and 3) applying a random frequency-shaping equalizer. As these scaling and frequency-shaping operations are uniformly applied across the utterance, they primarily modify the speaker's voice timbre while having minimal impact on the spoken content or accent. For each sample, a random variable $\gamma$ is drawn from the uniform distribution $\mathcal{U}(0, 1)$. Perturbations are applied if $\gamma > 0.3$; otherwise, the original waveforms are used. 
We show the effect of the data augmentation in the ablation study (Section. \ref{sec:ablation}).

\subsection{Inference-Time Steering for Accent Neutralization}
During inference, we apply the steering vectors to the corresponding layers of the backbone LLM: at each decoding step $t$, we modify the activations of layer $l$ as follows:
\begin{align}
    \mathbf{a}_{l}^{t} \leftarrow \left(\mathbf{a}_{l}^{t} - \alpha \cdot \mathbf{v}_l\right) \cdot \frac{||\mathbf{a}_{l}^{t}||_2}{||\mathbf{a}_{l}^{t} - \alpha \cdot \mathbf{v}_l||_2}
\end{align}
where $\mathbf{a}_{l}^{t} \in \mathbb{R}^{d}$ is the activation at layer $l$ and decoding step $t$, and $\alpha$ is a hyperparameter that controls the steering strength.
When the inference reference samples are accented, subtracting the steering vectors enables the negation of such ``accent directions'', which steers the accent activations towards neutral activations and thus mitigates the accent in the generated speech. 
The normalization term is applied to maintain the original activation norm. 
The steering vectors are only applied to the generated tokens, while the activations of the prompt tokens are not modified. In this study, we experiment with single-layer steering, where only one layer is steered while other layers are left unchanged.

\begin{table*}[th]
  \caption{Evaluation results on L2-ARCTIC and speechocean762. The steering vectors are extracted using the samples from L2-ARCTIC, and applied to both L2-ARCTIC and speechocean762 for evaluation. 
  EN\_US (US American English) and EN\_CN (Mandarin Chinese accented English) denote the reference speech accents; ISR denotes Inference Success Rate; AMR-CN and AMR-US denotes Accent Match Rate with EN\_CN and EN\_US accent target, respectively. See Section. \ref{sec:exp_setup} for more details on the evaluation metrics. Steering strength $\alpha$ is set to 1.0 for all the steered models in the table.}
  \label{tab:main}
  \centering
  \renewcommand{\arraystretch}{1.1} 
  \resizebox{\linewidth}{!}{%
  \begin{tabular}{@{} ccccccccccc @{}}
    \toprule
    \textbf{ID} & \makecell{\textbf{Evaluation} \\ \textbf{Dataset}} & \textbf{Model} & \makecell{\textbf{Steering} \\ \textbf{Setting}} & \makecell{\textbf{Prompt} \\ \textbf{Accent}} & \makecell{\textbf{ISR $\uparrow$} \\ \textbf{(\%)}} & \makecell{\textbf{AMR-CN $\downarrow$} \\ \textbf{(\%)}} & \makecell{\textbf{AMR-US $\uparrow$} \\ \textbf{(\%)}} & \makecell{\textbf{Spk Sim $\uparrow$}} & \textbf{UTMOS $\uparrow$} & \makecell{\textbf{WER $\downarrow$} \\ \textbf{(\%)}} \\ 
    \midrule
    1 & \multirow{8}{*}{L2-ARCTIC}                      
      & \multirow{4}{*}{Qwen3-TTS 0.6B} 
      & \multirow{2}{*}{Unsteered}    & EN\_US & 100.00            & 0.00             & 100.00            & 0.87          & 3.34          & 1.04          \\
    2 & &                             &               & EN\_CN & 98.46          & 82.14         & 0.00              & \textbf{0.85}          & \textbf{3.31}          & 3.98          \\ \cmidrule{5-11} 
    3 & &                             & Steer layer 15& EN\_CN & \textbf{98.68} & \textbf{1.78} & \textbf{97.33} & \textbf{0.73}          & \textbf{3.18}          & \textbf{2.64} \\
    4 & &                             & Steer layer 10& EN\_CN & 96.04          & 4.35          & 94.74          & 0.72          & 3.10           & 3.18          \\ \cmidrule{3-11} 
    5 & & \multirow{4}{*}{Qwen3-TTS 1.7B} 
      & \multirow{2}{*}{Unsteered}    & EN\_US & 100.00            & 0.00             & 100.00            & 0.87          & 3.38          & 0.97          \\
    6 & &                             &               & EN\_CN & \textbf{99.56} & 83.89         & 1.10            & \textbf{0.84} & \textbf{3.33} & 3.40           \\ \cmidrule{5-11} 
    7 & &                             & Steer layer 15& EN\_CN & \textbf{99.56} & \textbf{9.49}          & \textbf{88.74}          & \textbf{0.76} & \textbf{3.32} & \textbf{2.24} \\
    8 & &                             & Steer layer 10& EN\_CN & 99.34          & 18.14         & 79.87          & \textbf{0.76} & 3.22          & 2.63          \\ 
    \midrule
    \midrule
    9 & \multirow{3}{*}{speechocean762}                 
      & \multirow{3}{*}{Qwen3-TTS 1.7B} 
      & Unsteered                     & EN\_CN & \textbf{97.00}             & \textbf{3.09}          & 0.00              & \textbf{0.78}          & 2.73          & 56.41         \\ \cmidrule{5-11} 
    10& &                             & Steer layer 15& EN\_CN & \textbf{92.00}    & \textbf{3.26} & \textbf{48.91}          & \textbf{0.74} & \textbf{3.01}          & \textbf{32.43} \\
    11& &                             & Steer layer 10& EN\_CN & 86.00             & 6.98          & 48.84 & \textbf{0.74} & 2.93 & 32.85         \\ 
    \bottomrule
  \end{tabular}
  }
\end{table*}

\section{Experimental Setup} \label{sec:exp_setup}

\noindent\textbf{Datasets} We use ARCTIC and L2-ARCTIC for steering vector extraction. We focus on neutralizing the Mandarin Chinese-accented English. Speech and transcripts from the 4 Chinese-L1 speakers in L2-ARCTIC and the 4 native speakers in ARCTIC are used. For each speaker, we hold out 10\% of the utterances as the evaluation set, and we sample from the remaining utterances to extract steering vectors.
As an out-of-distribution evaluation to test the generalizability of the steering vectors, we use speechocean762 \cite{zhang2021speechocean762}, which features 250 Mandarin Chinese-L1 speakers (adults and children) with diverse English proficiency levels. We randomly sample 100 utterances from the adult group of the test set, which are paired with another 100 different text transcripts in speechocean762 as the target texts for TTS inference.

\noindent\textbf{Steering Vector Extraction and Steering Setting} We experiment with two sizes of the pre-trained Qwen3-TTS models, 0.6B and 1.7B parameters. Both models have 28 Transformer layers in the backbone LLM. For each model, we extract steering vectors from the following layer indices (zero-indexed): ${1,5,10,15,20,25,27}$. We use 4,000 ARCTIC plus L2-ARCTIC samples for steering vector extraction (i.e. $N_a + N_n=4000$ in Eqn. \ref{eq:sv_extraction}). During inference, we experiment with two steering strengths $\alpha$: 1.0 and 2.0. 

\noindent\textbf{Evaluation Metrics} We evaluate the generated speech on five objective metrics. 
1) \textbf{Inference Success Rate (ISR)}: the percentage of samples generated without errors, measuring stability under steering\footnote{With larger $\alpha$, and in rare cases even for the unsteered baseline, the model can get stuck in the decoding loop due to a missing stop token. We consider such cases as failed inference.};
2) \textbf{Accent Match Rate (AMR)}: the percentage of samples classified as a given accent (e.g., EN\_CN, EN\_US) by a pre-trained accent classifier \cite{yang23v_interspeech}\footnote{AMR-CN and AMR-US do not sum to 1 because the classifier was trained on 7 accents (ARCTIC + L2-ARCTIC).};
3) \textbf{Spk Sim}: cosine similarity between speaker embeddings of the generated and reference speech, extracted with a pre-trained speaker encoder\footnote{\url{https://github.com/resemble-ai/Resemblyzer}};
4) \textbf{UTMOS}: naturalness Mean of Opinion Score (MOS) predicted by UTMOSv2 \cite{baba2024utmosv2};
5) \textbf{Word Error Rate (WER)}: we use Whisper-turbo \cite{radford2023robust} to transcribe the generated speech and compute WER against the target text.

For subjective evaluation, we collect three MOS (scale 1--5, higher is better): 
1) \textbf{Naturalness MOS (NMOS)}, rating overall naturalness; 
2) \textbf{Accent MOS (AMOS)}, where listeners are presented with pairs of reference prompt speech (EN\_CN accent) and (un)steered TTS output, and score how much the accent of the TTS output is neutralized towards a US American English accent; 
3) \textbf{Speaker Timbre MOS (SMOS)}, where the same pairs are rated for timbre similarity to the reference.

\section{Results}

\subsection{In-Domain and Cross-Domain Steering}
Table \ref{tab:main} shows the evaluation results on L2-ARCTIC and speechocean762.
The unsteered models (Rows 1 and 2, 5 and 6) show high AMR following the prompt accent, meaning that the generated speech inherits the accent from the reference speech.
For both the 0.6B and 1.7B models, for the EN\_CN prompt accent, activation steering significantly reduces AMR-CN and increases AMR-US, indicating effective accent neutralization.
However, we also observe a Spk Sim drop, suggesting a trade-off between accent neutralization and timbre preservation. The 1.7B model's Spk Sim drop (0.84 to 0.76 ) is less than the 0.6B model (Rows 6 and 7 vs. Rows 2 and 3). 
One reason of the Spk Sim drop may be due to the speaker encoder model's sensitivity to accent changes. Prior works on Emotional Voice Conversion \cite{du24_odyssey} showed that speaker embeddings could be distant even for the same speaker with different emotions. We hypothesize that this may also be the case for accent changes.
After listening to the speech generated with steering, we find that in some cases, there are some local pitch shifts and prosody changes that may contribute to the accent neutralization.
Nevertheless, the overall speaker identity is still largely preserved. We provide an in-depth analysis of what the raw cosine similarity drop measures in Section. \ref{sec:secs_drop}. 

In addition, UTMOS is also improved or maintained after steering. This indicates the high naturalness of the generated speech under steering. The WER improvement may be attributed to the reduced accent (and potentially mispronunciations), making the generated speech more intelligible \cite{yang22v_interspeech}. 
The steering vectors also show strong generalizability to unseen accented speakers in speechocean762, as evidenced by the significant improvement of AMR-US compared to the unsteered condition. Notably, WER is significantly reduced after steering (56.41\% to 32.43\%, Rows 9 and 10). This may be due to the fact that speechocean762 speakers have much more diverse proficiency levels, and the speech recordings contain more pronunciation errors and disfluencies than the L2-ARCTIC speakers, making the zero-shot voice cloning more challenging and producing less natural and intelligible speech. The proposed steering vectors may provide a mitigation solution for such challenging cases.

\begin{figure}
    \centering
    \includegraphics[width=\linewidth]{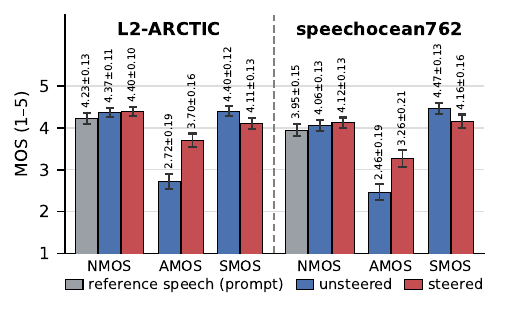}
    \vspace{-10mm}
    \caption{MOS and the 95\% confidence intervals of Naturalness (NMOS), Accent neutralization (AMOS) and Speaker timbre similarity (SMOS).}
    \label{fig:mos}
    \vspace{-3mm}
\end{figure}

Fig. \ref{fig:mos} shows the subjective evaluation results. We select 20 triplets of (reference prompt speech, unsteered TTS output, steered TTS output), and ask 13 human listeners to conduct the MOS evaluation (Section. \ref{sec:exp_setup}). The overall trends on L2-ARCTIC and speechocean762 are similar: compared to the unsteered group, the accent-steered group shows significantly higher AMOS, indicating a clear accent neutralization effect. Although the steered group shows a modest SMOS drop, the absolute SMOS is still high ($> 4.1$), meaning that the speaker timbre is largely preserved. In addition, both the unsteered and steered groups have high NMOS, even surpassing the reference group. This suggests that the proposed activation steering approach still produces highly natural TTS output.

\begin{figure*}[t]
  \centering
  \includegraphics[width=\linewidth]{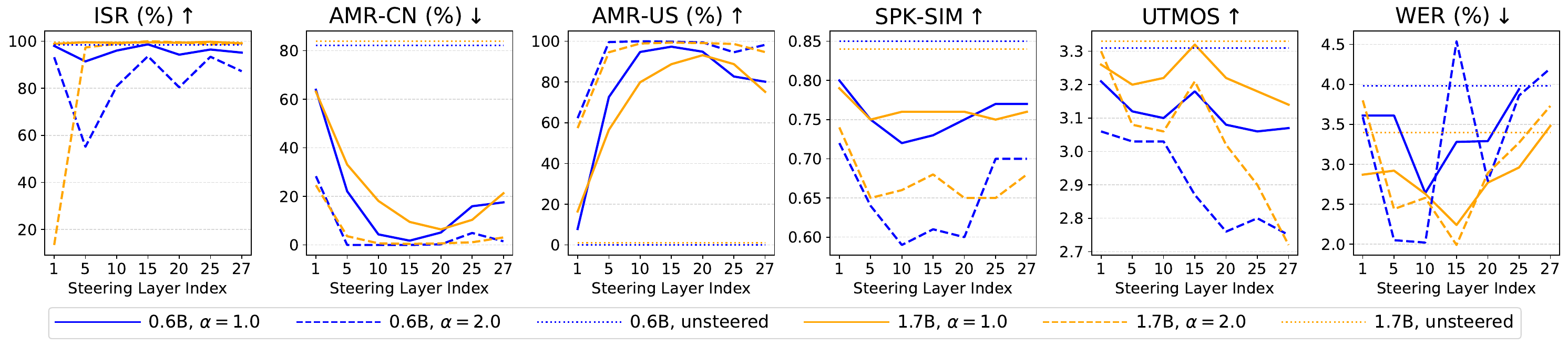}
  \vspace{-7mm}
  \caption{Layerwise single-layer steering analyses on L2-ARCTIC. Layers are zero-indexed (layer 1 means the 2nd layer). Solid and dashed lines represent the results of steering with different models and steering strengths. Dotted lines represent the unsteered baseline.}
  \label{fig:layerwise}
  \vspace{-3mm}
\end{figure*}

\subsection{Layerwise Steering Analysis}
We analyze the steering effect across layers.
Fig. \ref{fig:layerwise} shows the 6 metrics when steering these layers with different models and steering strengths. 
When $\alpha=1.0$, from the AMR-CN, AMR-US and Spk Sim curves, we can see that in general, steering the middle layers (15 and 20) provides the most balanced trade-off between accent neutralization and timbre preservation, while steering the early layers and top layers has a weaker effect on accent neutralization but stronger timbre preservation.
This echoes the findings in Persona Vectors \cite{chen2025persona}, which showed that steering middle layers has a stronger effect on persona trait expression. 
When increasing to $\alpha=2.0$, although the overall AMR-CN reduction is larger, Spk Sim also drops significantly.
This suggests a negative effect of over-steering the ``accent vectors": it aggressively deviates the representations far from the reference speaker's original activation space, causing drastic speaker identity change.

The UTMOS and WER curves show that steering middle layers (especially layer 15) also leads to better naturalness and intelligibility, even matching or surpassing the unsteered baseline, while steering early layers and top layers could cause naturalness and intelligibility degradation. When $\alpha=2.0$, steering early layers causes significant ISR drops, suggesting that early layers are more sensitive to over-steering and more likely to cause inference failure.
In addition, the 1.7B model shows better stability (steerability) than the 0.6B model, as evidenced by the higher ISR under the same steering settings.

\subsection{What Does the Cosine Similarity Drop Measure?} \label{sec:secs_drop}
Our main results (Table~\ref{tab:main}) report \textbf{Spk Sim} as the cosine similarity between Resemblyzer embeddings of the reference and generated speech, which falls from $0.84$ (unsteered) to $0.76$ under steering. We ask whether this drop is partly an artifact of the metric reacting to the intended accent change, and how much it affects the speaker-identity judgement. Formalizing accent-invariant measurements in an estimated accent-orthogonal subspace, we show that neutralized accent accounts for $\sim25\%$ of the drop, and that $>95\%$ of accent-neutralized utterances still verify as the reference speaker.

To estimate the accent-orthogonal subspace, we build a probe set: for each of the 28 ARCTIC and L2-ARCTIC speakers, we sample 150 utterances (disjoint from the steering test utterances) and apply the same timbre augmentation as in \ref{sec:sv_extraction} (4 augmented waveforms per utterance) to broaden per-accent timbre coverage. We then compute L2-normalized speaker embeddings $e$ for all original and augmented utterances. Let $\mu_a=\mathbb{E}[e\mid \text{accent}=a]$ be the centroid of accent $a$ and $\bar\mu=\frac1K\sum_a\mu_a$ the global centroid over $K$ accents. The matrix $A\in\mathbb{R}^{K\times d}$ stacks the centroid differences $A_a=\mu_a-\bar\mu$; since its rows sum to zero, $\text{Rank}(A)\leq K-1=6$. Its compact Singular Value Decomposition (SVD),
\begin{equation}
    A=U\Sigma V^\top
\end{equation}
with $U\in\mathbb R^{K\times (K-1)}$, $\Sigma\in \mathbb R^{(K-1)\times (K-1)}$, $V\in \mathbb R^{d\times (K-1)}$, has right singular vectors $v_1,\dots,v_{K-1}\in\mathbb R^{d}$ that span the accent subspace. The projector $P=I-VV^\top$ then maps $e$ to the orthogonal complement of this subspace:
\begin{equation}
    e^{\mathrm{accent-orthogonal}} = e - VV^\top e = (I-VV^\top)e
    \label{eqn:accent_ortho}
\end{equation}
We hypothesize that this subspace is more accent-invariant and better suited to evaluating speaker properties such as timbre independently of accent.

\begin{table}[t]
  \centering
  \caption{Accent and timbre probing results (Section. \ref{sec:secs_drop}). \textit{o-\textless encoder\textgreater} denotes the accent-orthogonalized speaker encoder.}
  \label{tab:probe}
  \begin{tabular}{cccc}
    \toprule
    \textbf{Speaker Encoder} & \textbf{Accent F1 $\downarrow$ (\%)}  & \textbf{TimbreR $\uparrow$}  & \textbf{EER $\downarrow$ (\%)} \\
    \midrule
    Resemb   & \textbf{28.96} & 3.95 & 3.35  \\
    ECAPA    & 34.57 & \textbf{5.87} & \textbf{0.75}  \\
    WavLM    & 40.02 & 2.21 & 9.73  \\
    \midrule
    o-Resemb & \textbf{0.00}  & 3.94 & 3.59  \\
    o-ECAPA  & \textbf{0.00}  & \textbf{6.35} & \textbf{0.48}  \\
    o-WavLM  & 0.54  & 2.23 & 16.55 \\
    \bottomrule
  \end{tabular}
\end{table}

\begin{table*}[t]
  \centering
  \caption{Re-scoring speaker timbre similarity in the accent-orthogonal subspace. Speaker Verification Acceptance Rate (SV-accept) is defined in Eqn. \ref{eqn:svaccept}. The evaluation is conducted on L2-ARCTIC with the EN\_CN prompt accent. Steering strength $\alpha$ is set to 1.0.}
  \label{tab:rescore}
  \resizebox{\textwidth}{!}{%
  \begin{tabular}{cccccccccccccc}
    \toprule
    \multirow{2}{*}[-0.6ex]{\makecell{\textbf{Model}}} & \multirow{2}{*}[-0.6ex]{\makecell{\textbf{Steering}}} & \multirow{2}{*}[-0.6ex]{\makecell{\textbf{AMR-CN $\downarrow$}\\\textbf{(\%)}}} & \multirow{2}{*}[-0.6ex]{\makecell{\textbf{AMR-US $\uparrow$}\\\textbf{(\%)}}} & \multirow{2}{*}[-0.6ex]{\textbf{UTMOS $\uparrow$}} & \multirow{2}{*}[-0.6ex]{\makecell{\textbf{WER $\downarrow$}\\\textbf{(\%)}}} & \multicolumn{4}{c}{\textbf{Speaker Embedding Cosine Similarity $\uparrow$}} & \multicolumn{4}{c}{\textbf{Speaker Verification Acceptance Rate $\uparrow$ (\%)}} \\
    \cmidrule(lr){7-10} \cmidrule(lr){11-14}
    & & & & & & \textbf{Resemb} & \textbf{o-Resemb} & \textbf{ECAPA} & \textbf{o-ECAPA} & \textbf{Resemb} & \textbf{o-Resemb} & \textbf{ECAPA} & \textbf{o-ECAPA} \\
    \midrule
    \multirow{2}{*}{Qwen3-TTS 0.6B} & Unsteered & 82.14 & 0.00  & 3.31 & 3.98 & 0.85 & 0.84 & 0.69 & 0.62 & 99.33 & 100.00 & 99.33 & 99.61 \\
                          & Steer L15 & 1.78  & 97.33 & 3.18 & 2.64 & 0.73 & 0.76 & 0.41 & 0.40 & 73.12 & 96.74  & 65.04 & 88.23 \\
    \midrule
    \multirow{2}{*}{Qwen3-TTS 1.7B} & Unsteered & 83.89 & 1.10  & 3.33 & 3.40 & 0.84 & 0.84 & 0.70 & 0.63 & 98.04 & 99.82  & 99.13 & 99.31 \\
                          & Steer L15 & 9.49  & 88.74 & 3.32 & 2.24 & 0.76 & 0.78 & 0.52 & 0.50 & 84.34 & 99.13  & 88.12 & 95.42 \\
    \bottomrule
  \end{tabular}%
  }
\end{table*}

We run the following probing experiments on the speaker embeddings to quantify accent decodability, timbre resolution, and speaker verification performance:

\noindent\textbf{Accent F1}. We train a linear Logistic Regression probe on the probe set (original and augmented waveforms) for 7-class accent classification, evaluated by 4-fold cross-validation: each fold holds out one speaker per accent as the test set (original waveforms only) and trains on the rest, so every utterance is predicted once. We report the macro-F1.

\noindent\textbf{Speaker Timbre Resolution (TimbreR)}. We formulate an accent-controlled identity-resolution measure on the probe set's original natural waveforms only. Sampling 300k random index pairs $(i,j),\,i\neq j$, we split their cosine similarities $c_{ij}$ into a genuine group $\mathcal{G}=\{c_{ij} | \mathrm{spk}_i=\mathrm{spk}_j\}$ (same speaker) and an impostor group $\mathcal{I}=\{c_{ij} | \mathrm{spk}_i\neq \mathrm{spk}_j,\ \mathrm{accent}_i=\mathrm{accent}_j\}$ (different speaker, \emph{same accent}). Restricting impostors to same-accent pairs isolates timbre, the identity signal that survives when accent is held constant. With genuine/impostor means and variances $(\mu_\mathcal{G},\sigma_\mathcal{G}^2)$, $(\mu_\mathcal{I},\sigma_\mathcal{I}^2)$:
\begin{equation}
    \mathrm{TimbreR}=\dfrac{\mu_\mathcal{G}-\mu_\mathcal{I}}{\sqrt{\tfrac12(\sigma_\mathcal{G}^2+\sigma_\mathcal{I}^2)}}
    \label{eqn:timbrer}
\end{equation}
estimates the standardized separability of timbre.

\noindent\textbf{Speaker Verification Equal Error Rate (EER)}. Following a Speaker Verification (SV) process, we decide whether two utterances share a speaker by thresholding their cosine similarity. For threshold $t$, the False Reject and False Accept Rates are $\mathrm{FRR}(t)=|\{c\in\mathcal G | c < t\}|/ |\mathcal G|$ and $\mathrm{FAR}(t)=|\{c\in\mathcal I | c \ge t\}|/ |\mathcal I|$. The threshold $t^\star$ minimizing $|\mathrm{FRR}-\mathrm{FAR}|$ gives the EER:
$$t^\star=\arg\min_t|\mathrm{FRR}(t)-\mathrm{FAR}(t)|$$
\begin{equation}
    \mathrm{EER}=\tfrac12\big(\mathrm{FRR}(t^\star)+\mathrm{FAR}(t^\star)\big)
    \label{eqn:EER}
\end{equation}
We obtain $t^\star$ from the probe set, excluding the steering test speakers to prevent leakage, and apply it to the steered outputs to compute the Speaker Verification Acceptance Rate (SV-accept): for each utterance $u$ in the steering test set $\mathcal D$, let $c_u=\cos(u^{ref}, u^{gen})$ be the cosine similarity between the reference speech and the TTS-generated speech (steered or unsteered), then
\begin{equation}
    \text{SV-accept} = \frac{|\{u \in \mathcal D | c_u \ge t^\star \}|}{|\mathcal D|}
    \label{eqn:svaccept}
\end{equation}
is the fraction of TTS outputs an SV system verifies as the reference speaker.

Table~\ref{tab:probe} reports the Accent F1, TimbreR (Eqn.~\ref{eqn:timbrer}), and EER (Eqn.~\ref{eqn:EER}) for Resemblyzer and two additional encoders, SpeechBrain ECAPA-TDNN\footnote{\url{https://huggingface.co/speechbrain/spkrec-ecapa-voxceleb}} \cite{ecapa, speechbrain} and WavLM-base-plus-sv\footnote{\url{https://huggingface.co/microsoft/wavlm-base-plus-sv}} \cite{wavlm}, each with its accent-orthogonalized variant (Eqn.~\ref{eqn:accent_ortho}), denoted as \textit{o-\textless encoder\textgreater}. We want accent-invariant (low Accent F1) embeddings that assess timbre more reliably. All three raw encoders are accent-leaky (above the $1/7$ chance F1), with Resemblyzer being the most accent-invariant, while ECAPA-TDNN is the most timbre-discriminative with the lowest EER. Orthogonalization drives Accent F1 to near $0$ for all (``accent erasure''). It even improves ECAPA-TDNN's resolution (TimbreR: 5.87$\to$6.35, EER: 0.75\%$\to$0.48\%), while barely affecting Resemblyzer, and significantly degrading WavLM (EER: 9.73\%$\to$16.55\%). Thus Resemblyzer and ECAPA-TDNN have a more separable accent subspace: erasing it leaves timbre and overall identity intact or improved. WavLM encodes these properties more non-linearly, so linear erasure harms verification, making it unsuitable for evaluating accent-neutralized speaker similarity here.

Guided by these findings, we rescore speaker timbre similarity of the steering experiments (Table \ref{tab:main}) in the accent-orthogonal subspace using Resemblyzer and ECAPA-TDNN, adding SV-accept (Eqn.~\ref{eqn:svaccept}) as a further metric (Table~\ref{tab:rescore}). For the 1.7B model, orthogonalization shrinks the cosine-similarity drop (Resemblyzer: 0.08$\to$0.06, ECAPA-TDNN: 0.18$\to$0.13), suggesting $\sim$25\% of the drop is an accent artifact. The original-space SV-accept (Resemblyzer: 84.34\%, ECAPA-TDNN: 88.12\%) conflate accent change with other voice characteristics differences. In the accent-orthogonal subspace, they rise to 99.13\% and 95.42\% respectively. Hence, once accent is ruled out, over 95\% of neutralized utterances still verify as the reference speaker, indicating that the raw cosine similarities in Table~\ref{tab:main} overstate the identity change from accent.
Corroborating with the subjective evaluation in Fig. \ref{fig:mos}, both subjective evaluations and our accent-orthogonal objective evaluations confirm that speaker timbre is largely preserved.

\begin{table}[tbp]
  \caption{Ablation study: impact of the number of samples and data augmentation for steering vectors extraction (Section. \ref{sec:sv_extraction}). The steering vectors are extracted from the 1.7B model's layer 15. The evaluation is conducted on L2-ARCTIC with EN\_CN prompt accent. Steering strength $\alpha$ is set to 1.0.} 
  \label{tab:aug} 
  \centering
  \renewcommand{\arraystretch}{1.1} 
  \resizebox{\linewidth}{!}{%
  \begin{tabular}{@{} ccccccc @{}}
    \toprule
    \makecell{\textbf{\# Samples}} & \textbf{Augmentation} & \makecell{\textbf{ISR$\uparrow$} \\ \textbf{(\%)}} & \makecell{\textbf{AMR-US$\uparrow$} \\ \textbf{(\%)}} & \makecell{\textbf{Spk Sim$\uparrow$}} & \textbf{UTMOS$\uparrow$} & \makecell{\textbf{WER$\downarrow$} \\ \textbf{(\%)}} \\ 
    \midrule
    \multirow{2}{*}{4000} & w/ aug.  & 99.56                 & 88.74          & \textbf{0.76} & \textbf{3.32} & \textbf{2.24} \\
                          & w/o aug. & \textbf{100.00}   & \textbf{93.41} & 0.75          & 3.27          & 2.49          \\ 
    \midrule
    \multirow{2}{*}{1000} & w/ aug.  & 99.56                 & 87.42          & \textbf{0.77} & \textbf{3.32} & 2.42          \\
                          & w/o aug. & \textbf{99.78}& \textbf{93.61} & 0.74          & 3.27          & \textbf{2.24} \\ 
    \midrule
    \multirow{2}{*}{40}   & w/ aug.  & \textbf{100.00}            & 93.19          & \textbf{0.73} & \textbf{3.28} & \textbf{2.14} \\
                          & w/o aug. & \textbf{100.00}   & \textbf{95.60}  & 0.70           & 3.27          & 2.46          \\ 
    \bottomrule
  \end{tabular}%
  }
\end{table}

\subsection{Ablation Study on Steering Vectors Extraction} \label{sec:ablation}
In this section, we investigate the impact of the number of samples and data augmentation for steering vector extraction (Section. \ref{sec:sv_extraction}). We experiment with three different numbers of samples: 40, 1000, and 4000. For each setting, we compare the steering vectors extracted with and without data augmentation. 
The results are shown in Table \ref{tab:aug}. We find that the data augmentation effectively improves Spk Sim. This suggests that the proposed data augmentation can help break the entanglement between accent and speaker identity, and encourage the steering vectors to capture more accent-specific information.
In terms of the number of needed samples for steering vector extraction, it seems that 1000 samples are sufficient to achieve a decent balance between accent neutralization and timbre preservation.

\section{Conclusion}
In this work, we introduce a novel, simple yet effective activation steering method for accent-neutralized zero-shot TTS. 
We extract layer-wise steering vectors based on the activation differences between accented and neutral speech. 
During inference, these vectors guide the model toward generating accent-neutralized speech. 
Experimental results on Qwen3-TTS demonstrate that the proposed steering vectors effectively reduce accents in the synthesized speech while largely retaining the speaker's timbre.
Furthermore, the steering vectors exhibit strong generalizability to unseen accented speakers, indicating that they capture a universal direction for accent neutralization within the activation space of the TTS model.
Our analysis reveals that raw cosine similarities conflate accent change with other voice characteristics. We formulate objective measurements in an estimated accent-orthogonal subspace, which provide a more faithful way to evaluate speaker timbre similarity.

\newpage
\section{Generative AI Use Disclosure}
The authors used Google's Gemini and Anthropic's Claude to polish and edit the manuscript. The authors additionally used Claude Code to write portions of the code used for the probing experiments and analyses reported in Section. \ref{sec:secs_drop}. All research ideas, experimental designs, experiments, analyses, and conclusions were conceived, carried out, and validated by the authors. The authors reviewed and verified all AI-assisted outputs, including all generated code, and take full responsibility for the content of this article.

\bibliographystyle{IEEEtran}
\bibliography{mybib}

@inproceedings{chen2025f5,
  author = {Chen, Yushen and Niu, Zhikang and Ma, Ziyang and Deng, Keqi and Wang, Chunhui and JianZhao, JianZhao and Yu, Kai and Chen, Xie},
  title = {{F5-TTS}: A fairytaler that fakes fluent and faithful speech with flow matching},
  booktitle = {Proc. Annual Meeting of the Association for Computational Linguistics (ACL), Volume 1: Long Papers},
  pages = {6255--6271},
  year = {2025}
}

@inproceedings{eskimez2024e2,
  author = {Eskimez, Sefik Emre and Wang, Xiaofei and Thakker, Manthan and Li, Canrun and Tsai, Chung-Hsien and Xiao, Zhen and Yang, Hemin and Zhu, Zirun and Tang, Min and Tan, Xu and others},
  title = {{E2 TTS}: Embarrassingly easy fully non-autoregressive zero-shot {TTS}},
  booktitle = {Proc. IEEE Spoken Language Technology Workshop (SLT)},
  pages = {682--689},
  year = {2024},
  organization = {IEEE}
}

@inproceedings{kim2023p,
  author = {Kim, Sungwon and Shih, Kevin and Santos, Joao Felipe and Bakhturina, Evelina and Desta, Mikyas and Valle, Rafael and Yoon, Sungroh and Catanzaro, Bryan and others},
  title = {{P-Flow}: A fast and data-efficient zero-shot {TTS} through speech prompting},
  booktitle = {Proc. Advances in Neural Information Processing Systems (NeurIPS)},
  volume = {36},
  pages = {74213--74228},
  year = {2023}
}

@inproceedings{Ju2024naturalspeech,
  author = {Ju, Zeqian and Wang, Yuancheng and Shen, Kai and Tan, Xu and Xin, Detai and Yang, Dongchao and Liu, Yanqing and Leng, Yichong and Song, Kaitao and Tang, Siliang and Wu, Zhizheng and Qin, Tao and Li, Xiang-Yang and Ye, Wei and Zhang, Shikun and Bian, Jiang and He, Lei and Li, Jinyu and Zhao, Sheng},
  title = {{NaturalSpeech 3}: Zero-shot speech synthesis with factorized codec and diffusion models},
  booktitle = {Proc. International Conference on Machine Learning (ICML)},
  year = {2024},
  publisher = {JMLR.org},
  articleno = {909},
  numpages = {19},
  location = {Vienna, Austria}
}

@inproceedings{zhu2025zipvoice,
  author = {Zhu, Han and Kang, Wei and Yao, Zengwei and Guo, Liyong and Kuang, Fangjun and Li, Zhaoqing and Zhuang, Weiji and Lin, Long and Povey, Daniel},
  title = {{ZipVoice}: Fast and high-quality zero-shot text-to-speech with flow matching},
  booktitle = {Proc. IEEE Automatic Speech Recognition and Understanding Workshop (ASRU)},
  year = {2025}
}

@article{hu2026qwen3,
  author = {Hu, Hangrui and Zhu, Xinfa and He, Ting and Guo, Dake and Zhang, Bin and Wang, Xiong and Guo, Zhifang and Jiang, Ziyue and Hao, Hongkun and Guo, Zishan and others},
  title = {{Qwen3-TTS} technical report},
  journal = {arXiv preprint arXiv:2601.15621},
  year = {2026}
}

@article{wang2025spark,
  author = {Wang, Xinsheng and Jiang, Mingqi and Ma, Ziyang and Zhang, Ziyu and Liu, Songxiang and Li, Linqin and Liang, Zheng and Zheng, Qixi and Wang, Rui and Feng, Xiaoqin and others},
  title = {{Spark-TTS}: An efficient {LLM}-based text-to-speech model with single-stream decoupled speech tokens},
  journal = {arXiv preprint arXiv:2503.01710},
  year = {2025}
}

@article{ye2025llasa,
  author = {Ye, Zhen and Zhu, Xinfa and Chan, Chi-Min and Wang, Xinsheng and Tan, Xu and Lei, Jiahe and Peng, Yi and Liu, Haohe and Jin, Yizhu and Dai, Zheqi and others},
  title = {{Llasa}: Scaling train-time and inference-time compute for {Llama}-based speech synthesis},
  journal = {arXiv preprint arXiv:2502.04128},
  year = {2025}
}

@article{du2024cosyvoice1,
  author = {Du, Zhihao and Chen, Qian and Zhang, Shiliang and Hu, Kai and Lu, Heng and Yang, Yexin and Hu, Hangrui and Zheng, Siqi and Gu, Yue and Ma, Ziyang and others},
  title = {{CosyVoice}: A scalable multilingual zero-shot text-to-speech synthesizer based on supervised semantic tokens},
  journal = {arXiv preprint arXiv:2407.05407},
  year = {2024}
}

@article{du2024cosyvoice2,
  author = {Du, Zhihao and Wang, Yuxuan and Chen, Qian and Shi, Xian and Lv, Xiang and Zhao, Tianyu and Gao, Zhifu and Yang, Yexin and Gao, Changfeng and Wang, Hui and others},
  title = {{CosyVoice 2}: Scalable streaming speech synthesis with large language models},
  journal = {arXiv preprint arXiv:2412.10117},
  year = {2024}
}

@article{guo2025fireredtts,
  author = {Guo, Hao-Han and Hu, Yao and Shen, Fei-Yu and Tang, Xu and Wu, Yi-Chen and Xie, Feng-Long and Xie, Kun},
  title = {{FireRedTTS-1S}: An upgraded streamable foundation text-to-speech system},
  journal = {arXiv preprint arXiv:2503.20499},
  year = {2025}
}

@article{zhou2025indextts2,
  author = {Zhou, Siyi and Zhou, Yiquan and He, Yi and Zhou, Xun and Wang, Jinchao and Deng, Wei and Shu, Jingchen},
  title = {{IndexTTS2}: A breakthrough in emotionally expressive and duration-controlled auto-regressive zero-shot text-to-speech},
  journal = {arXiv preprint arXiv:2506.21619},
  year = {2025}
}

@inproceedings{zhangvevo,
  author = {Zhang, Xueyao and Zhang, Xiaohui and Peng, Kainan and Tang, Zhenyu and Manohar, Vimal and Liu, Yingru and Hwang, Jeff and Li, Dangna and Wang, Yuhao and Chan, Julian and others},
  title = {{Vevo}: Controllable zero-shot voice imitation with self-supervised disentanglement},
  booktitle = {Proc. International Conference on Learning Representations (ICLR)},
  year = {2025}
}

@article{hirschi2025artificial,
  author = {Hirschi, Kevin and Kang, Okim and Yang, Mu and Hansen, John HL and Beloin, Kyle},
  title = {Artificial intelligence-generated feedback for second language intelligibility: An exploratory intervention study on effects and perceptions},
  journal = {Language Learning},
  volume = {75},
  number = {S1},
  pages = {204--241},
  year = {2025},
  publisher = {Wiley Online Library}
}

@inproceedings{yang22v_interspeech,
  author = {Mu Yang and Kevin Hirschi and Stephen Daniel Looney and Okim Kang and John H.L. Hansen},
  title = {Improving mispronunciation detection with {Wav2vec2}-based momentum pseudo-labeling for accentedness and intelligibility assessment},
  booktitle = {Proc. Interspeech},
  pages = {4481--4485},
  year = {2022},
  doi = {10.21437/Interspeech.2022-11039},
  issn = {2958-1796}
}

@article{halychanskyi2025fac,
  author = {Halychanskyi, Yurii and Churchwell, Cameron and Wen, Yutong and Kindratenko, Volodymyr},
  title = {{FAC-FACodec}: Controllable zero-shot foreign accent conversion with factorized speech codec},
  journal = {arXiv preprint arXiv:2510.10785},
  year = {2025}
}

@inproceedings{quamer22_interspeech,
  author = {Waris Quamer and Anurag Das and John Levis and Evgeny Chukharev-Hudilainen and Ricardo Gutierrez-Osuna},
  title = {Zero-shot foreign accent conversion without a native reference},
  booktitle = {Proc. Interspeech},
  pages = {4920--4924},
  year = {2022},
  doi = {10.21437/Interspeech.2022-10664},
  issn = {2958-1796}
}

@inproceedings{gaintseva2026casteer,
  author = {Tatiana Gaintseva and Andreea-Maria Oncescu and Chengcheng Ma and Ziquan Liu and Martin Benning and Gregory Slabaugh and Jiankang Deng and Ismail Elezi},
  title = {{CASteer}: Cross-attention steering for controllable concept erasure},
  booktitle = {Proc. International Conference on Learning Representations (ICLR)},
  year = {2026},
  url = {https://openreview.net/forum?id=6D5Odqol1B}
}

@article{turner2023steering,
  author = {Turner, Alexander Matt and Thiergart, Lisa and Leech, Gavin and Udell, David and Vazquez, Juan J and Mini, Ulisse and MacDiarmid, Monte},
  title = {Steering language models with activation engineering},
  journal = {arXiv preprint arXiv:2308.10248},
  year = {2023}
}

@article{chen2025persona,
  author = {Chen, Runjin and Arditi, Andy and Sleight, Henry and Evans, Owain and Lindsey, Jack},
  title = {Persona vectors: Monitoring and controlling character traits in language models},
  journal = {arXiv preprint arXiv:2507.21509},
  year = {2025}
}

@inproceedings{suni25_ssw,
  author = {Antti Suni and Sébastien {Le Maguer} and Sofoklis Kakouros and Tuukka Törö and Juraj Šimko},
  title = {Style and prosody control for zero-shot speech synthesis},
  booktitle = {Proc. Speech Synthesis Workshop (SSW)},
  pages = {28--34},
  year = {2025},
  doi = {10.21437/SSW.2025-5}
}

@inproceedings{lee25f_interspeech,
  author = {Kyowoon Lee and Artyom Stitsyuk and Gunu Jho and Inchul Hwang and Jaesik Choi},
  title = {Counterfactual activation editing for post-hoc prosody and mispronunciation correction in {TTS} models},
  booktitle = {Proc. Interspeech},
  pages = {434--438},
  year = {2025},
  doi = {10.21437/Interspeech.2025-723},
  issn = {2958-1796}
}

@article{zhou2026emoshift,
  author = {Zhou, Li and Jiang, Hao and Li, Junjie and Wang, Tianrui and Li, Haizhou},
  title = {{EmoShift}: Lightweight activation steering for enhanced emotion-aware speech synthesis},
  journal = {arXiv preprint arXiv:2601.22873},
  year = {2026}
}

@article{lee2026erasing,
  author = {Lee, Myungjin and Shin, Eunji and Lee, Jiyoung},
  title = {Erasing your voice before it's heard: Training-free speaker unlearning for zero-shot text-to-speech},
  journal = {arXiv preprint arXiv:2601.20481},
  year = {2026}
}

@article{xie2025emosteer,
  author = {Xie, Tianxin and Yang, Shan and Li, Chenxing and Yu, Dong and Liu, Li},
  title = {{EmoSteer-TTS}: Fine-grained and training-free emotion-controllable text-to-speech via activation steering},
  journal = {arXiv preprint arXiv:2508.03543},
  year = {2025}
}

@article{defossez2024moshi,
  author = {D{\'e}fossez, Alexandre and Mazar{\'e}, Laurent and Orsini, Manu and Royer, Am{\'e}lie and P{\'e}rez, Patrick and J{\'e}gou, Herv{\'e} and Grave, Edouard and Zeghidour, Neil},
  title = {{Moshi}: A speech-text foundation model for real-time dialogue},
  journal = {arXiv preprint arXiv:2410.00037},
  year = {2024}
}

@inproceedings{kominek2004cmu,
  author = {Kominek, John and Black, Alan W},
  title = {The {CMU Arctic} speech databases},
  booktitle = {Proc. ISCA Speech Synthesis Workshop (SSW)},
  year = {2004}
}

@inproceedings{zhao2018l2,
  author = {Zhao, Guanlong and Sonsaat, Sinem and Silpachai, Alif and others},
  title = {{L2-ARCTIC}: A non-native {English} speech corpus},
  booktitle = {Proc. Interspeech},
  year = {2018}
}

@inproceedings{qian2022contentvec,
  author = {Qian, Kaizhi and Zhang, Yang and Gao, Heting and Ni, Junrui and Lai, Cheng-I and Cox, David and Hasegawa-Johnson, Mark and Chang, Shiyu},
  title = {{ContentVec}: An improved self-supervised speech representation by disentangling speakers},
  booktitle = {Proc. International Conference on Machine Learning (ICML)},
  pages = {18003--18017},
  year = {2022},
  organization = {PMLR}
}

@inproceedings{yang23v_interspeech,
  author = {Mu Yang and Ram C. M. C. Shekar and Okim Kang and John H. L. Hansen},
  title = {What can an accent identifier learn? {Probing} phonetic and prosodic information in a {Wav2vec2}-based accent identification model},
  booktitle = {Proc. Interspeech},
  pages = {1923--1927},
  year = {2023},
  doi = {10.21437/Interspeech.2023-2254},
  issn = {2958-1796}
}

@inproceedings{zhang2021speechocean762,
  author = {Zhang, Junbo and Zhang, Zhiwen and Wang, Yongqing and Yan, Zhiyong and Song, Qiong and Huang, Yukai and Li, Ke and Povey, Daniel and Wang, Yujun},
  title = {{speechocean762}: An open-source non-native {English} speech corpus for pronunciation assessment},
  booktitle = {Proc. Interspeech},
  year = {2021}
}

@inproceedings{baba2024utmosv2,
  author = {Baba, Kaito and Nakata, Wataru and Saito, Yuki and Saruwatari, Hiroshi},
  title = {The {T05} system for the {VoiceMOS Challenge 2024}: Transfer learning from deep image classifier to naturalness {MOS} prediction of high-quality synthetic speech},
  booktitle = {Proc. IEEE Spoken Language Technology Workshop (SLT)},
  pages = {818--824},
  year = {2024},
  doi = {10.1109/SLT61566.2024.10832315}
}

@inproceedings{radford2023robust,
  author = {Radford, Alec and Kim, Jong Wook and Xu, Tao and Brockman, Greg and McLeavey, Christine and Sutskever, Ilya},
  title = {Robust speech recognition via large-scale weak supervision},
  booktitle = {Proc. International Conference on Machine Learning (ICML)},
  pages = {28492--28518},
  year = {2023},
  organization = {PMLR}
}

@inproceedings{du24_odyssey,
  author = {Zongyang Du and Junchen Lu and Kun Zhou and Lakshmish Kaushik and Berrak Sisman},
  title = {Converting anyone's voice: End-to-end expressive voice conversion with a conditional diffusion model},
  booktitle = {Proc. Speaker and Language Recognition Workshop (Odyssey)},
  pages = {172--179},
  year = {2024},
  doi = {10.21437/odyssey.2024-25}
}

@inproceedings{li2023inference,
  author = {Li, Kenneth and Patel, Oam and Vi{\'e}gas, Fernanda and Pfister, Hanspeter and Wattenberg, Martin},
  title = {Inference-time intervention: Eliciting truthful answers from a language model},
  booktitle = {Proc. Advances in Neural Information Processing Systems (NeurIPS)},
  volume = {36},
  pages = {41451--41530},
  year = {2023}
}

@inproceedings{rodriguez2025controlling,
  author = {Pau Rodriguez and Arno Blaas and Michal Klein and Luca Zappella and Nicholas Apostoloff and Marco Cuturi and Xavier Suau},
  title = {Controlling language and diffusion models by transporting activations},
  booktitle = {Proc. International Conference on Learning Representations (ICLR)},
  year = {2025},
  url = {https://openreview.net/forum?id=l2zFn6TIQi}
}

@inproceedings{ecapa,
  author = {Brecht Desplanques and Jenthe Thienpondt and Kris Demuynck},
  title = {{ECAPA-TDNN}: Emphasized channel attention, propagation and aggregation in {TDNN} based speaker verification},
  booktitle = {Proc. Interspeech},
  pages = {3830--3834},
  year = {2020},
  publisher = {{ISCA}}
}

@article{speechbrain,
  author = {Mirco Ravanelli and Titouan Parcollet and Peter Plantinga and Aku Rouhe and Samuele Cornell and Loren Lugosch and Cem Subakan and Nauman Dawalatabad and Abdelwahab Heba and Jianyuan Zhong and Ju-Chieh Chou and Sung-Lin Yeh and Szu-Wei Fu and Chien-Feng Liao and Elena Rastorgueva and François Grondin and William Aris and Hwidong Na and Yan Gao and Renato De Mori and Yoshua Bengio},
  title = {{SpeechBrain}: A general-purpose speech toolkit},
  journal = {arXiv preprint arXiv:2106.04624},
  year = {2021}
}

@article{wavlm,
  author = {Chen, Sanyuan and Wang, Chengyi and Chen, Zhengyang and Wu, Yu and Liu, Shujie and Chen, Zhuo and Li, Jinyu and Kanda, Naoyuki and Yoshioka, Takuya and Xiao, Xiong and Wu, Jian and Zhou, Long and Ren, Shuo and Qian, Yanmin and Qian, Yao and Wu, Jian and Zeng, Michael and Yu, Xiangzhan and Wei, Furu},
  title = {{WavLM}: Large-scale self-supervised pre-training for full stack speech processing},
  journal = {IEEE Journal of Selected Topics in Signal Processing},
  volume = {16},
  number = {6},
  pages = {1505--1518},
  year = {2022},
  doi = {10.1109/JSTSP.2022.3188113}
}

@article{du2025cosyvoice,
  author = {Du, Zhihao and Gao, Changfeng and Wang, Yuxuan and Yu, Fan and Zhao, Tianyu and Wang, Hao and Lv, Xiang and Wang, Hui and Shi, Xian and An, Keyu and others},
  title = {{CosyVoice 3}: Towards in-the-wild speech generation via scaling-up and post-training},
  journal = {arXiv preprint arXiv:2505.17589},
  year = {2025}
}
\end{document}